\documentclass[submission,copyright,creativecommons]{eptcs}
\providecommand{\event}{WIVACE 2013} % Name of the event you are submitting to
\usepackage{breakurl}             % Not needed if you use pdflatex only.
\usepackage{color}         % produces boxes or entire pages with colored backgrounds
\usepackage{graphics}      % standard graphics specifications
\usepackage[pdftex]{graphicx}
\usepackage{epsf}          % old package handles encapsulated post script issues
\usepackage{bm}            % special 'bold-math' package
\usepackage{verbatim}			% for comment environment
\usepackage{subfigure}
\usepackage{hyperref}
\usepackage{amsmath}

\title{Impact of local information in growing networks}
\author{Emanuele Massaro$^*$, Henrik Olsson$^\S \ $, Andrea Guazzini$^\P \ $ and Franco Bagnoli$^\dagger \ $\\
\small{$^*$Dept. of Information engineering, University of Florence, Italy.}\\
\small{$^\S\ $Max Planck Institute for Human Development, Berlin, Germany.}\\
\small{$^\P \ $Dept. of Psychology, University of Florence, Italy.}\\
\small{$^\dagger \ $Dept. of Physics and Astronomy, University of Florence. Also INFN, sez. Firenze.}}
\def\titlerunning{Impact of local information in growing networks}
\def\authorrunning{E. Massaro, H. Olsson, A. Guazzini \& F. Bagnoli}
\begin{document}
\maketitle

\begin{abstract}
We present a new model of the evolutionary dynamics and the growth of on-line social networks. The model emulates people's strategies for acquiring information in social networks, emphasising the local subjective view of an individual and what kind of information the individual can acquire when arriving in a new social context. The model proceeds through two phases: (a) a discovery phase, in which the individual becomes aware of the surrounding world and (b) an elaboration phase, in which the individual elaborates locally the information trough a cognitive-inspired algorithm. Model generated networks reproduce main features of both theoretical and real-world networks, such as high clustering coefficient, low characteristic path length, strong division in communities, and variability of degree distributions.
\end{abstract}

\section{Introduction}
The emergence and the global adaptation of social networks has influenced human interaction on individual, community, and larger social levels. Most notably, perhaps, is the rise of Facebook, which in October 2011 reached more than half ($55 \%$) of the world’s global audiences, catching 835.6 millions of users in 2012~\cite{facebookstats}. Understanding the growth and development of social networks is a task of great importance in many disciplines, such as sociology, biology, and computer science~\cite{Waaserman,Scott,Mendes,Strogatz,Reka2002}, where systems are often represented as graphs. A large number of models have been proposed that aim at exploring and explaining how local mechanisms of network formation produce global network structure. In the context of social networks it is important to understand why and how people decide to make connections and how they change or modify their own local structure. For this reason it is essential to understand some aspects of how humans behave in social networks: How do people acquire information in on-line social networks and what are the mechanisms that lead people to join together or to visit a specific website?
% Che vuol dire questa frase?
%When we get information, for instance by ``googling" or looking the friend of friend %Facebook page, we decide if link to it or not.

One of the the most well known mechanism that is used in growing networks is preferential attachment, where new connections are established preferentially to more popular nodes in a network, giving rise to a scale-free network~\cite{barabasi}.
Moreover, users in on-line social networks tend to form groups, called communities: given a graph, a community is a group of vertices ``more linked'' among them than between the group and the rest of the graph~\cite{Girvan02}. This is clearly a poor definition, and indeed, on a connected graph, there is no clear distinction between a community and the rest of the graph. In general, there is a continuum of nested communities whose boundaries are somewhat arbitrary: the structure of communities can be seen as a hierarchical dendrogram~\cite{Newman}.  Our communities are large and varied, and we recognize several levels of grouping, sometimes dependent on the context. In recent work we have shown that using information dynamics algorithms where nodes elaborate information locally, we are able to detect such communities in complex networks~\cite{Massaro2012, Bagnoli2012}.

Recently, Papadopoulos et al.~\cite{popularity} explored the trade-off between popularity and similarity in growing networks. Nodes in growing networks tend to link not only to the most popular nodes (as in preferential attachment~\cite{barabasi}) but also to the closest nodes in terms of affinity. Comparing their results with real-world complex networks, the authors showed that they were able to predict the probability of forming new links with remarkable precision.

In this paper we develop a model that emulates the growing of a social network, starting from psychological assumptions that allow us to simulate how people acquire and elaborate information in social networks. We demonstrate the concept of similarity and popularity in growing networks, not by a geometric approach as in Papadopoulos et al.~\cite{popularity}, but by using a simple mechanism that explain users' behaviour in on-line social networks.

The rest of this paper is organized as follows: we start by summarizing previous work in section~\ref{previouswork}. In section~\ref{model} we describe our model, which uses a local algorithm where an agent is modeled with a memory and a set of connections to other individuals. In the first step the new agent explore the local structure of the network, where it receives information about the neighborhood. The learning (nonlinear) phase is modeled after competition in the chemical/ecological world, where agents compete with each other.  Section~\ref{results} shows the principal results of our simulations. Finally, we discuss our results and propose future steps in the Conclusion.

\section{Related work}
\label{previouswork}

The goal of much of the research that model the growth of real networks is to reproduce networks with certain properties as well as properties of real-world networks. For instance, we know that many observed networks fall into the class of scale-free networks, meaning that they have power-law (or scale-free) degree distributions. An influential model is the so called Barab\'{a}si–Albert (BA) model~\cite{barabasi}. The main hypothesis of the BA model is that the more connected a node is, the more likely it is to receive new links. From the BA model, however, it is not trivial to generate networks with a community structure or with a high clustering coefficient, something we observe in real social networks. Regarding social networks, Jin et al.~\cite{jin01} presented a model where the friendship between individuals depends on the number of mutual friends and the number of meetings between them. The resulting networks from their simulations show high levels of clustering and a strong community structure in which individuals have more links to others within their community than to individuals from other communities.

In these two models we can already single out two important and psychologically plausible features: (1) the predisposition of people to link with hubs (nodes with higher connectivity degree, where the connectivity degree, $C(k)$, of a node in a unweighted network is defined as the sum of its links) and (2) the tendency of people to connect with friends of friends (social community).

A recent paper highlighted the relevance of existing communities in a network~\cite{Buscarino}. Here, a new node connects to a random node in the network and with a probability $p$ it links to some nodes in the community of the selected node and with probability $1-p$ with some random nodes. Simulation results showed that this model can reproduce features of real world networks, but the authors used a global community detection algorithm, the well-known Louvain Method~\cite{fast}, without considering the local and subjective view of agents in the network.

Popularity is not the only thing that determines if a node will link to another node, social closeness is also important. It has been shown that more similar nodes have higher chances to connect to each other even if they are not popular~\cite{Watts02}: this effect is known as \emph{homophily} in social science~\cite{Şimşek2008}. As mentioned above, in 2012 Papadopoulos et al.~\cite{popularity} published a paper giving an important contribution to the understanding of the evolutionary properties of complex networks. The authors showed that the popularity of a node in a network is just one dimension of attractiveness. In their framework, the probability that a new node that arrives in a structured network links to another node is a function of two variables: the connectivity degree of the target node (popularity) and the similarity (affinity) between the new node and the target node. In order to evaluate the trade-off between popularity and similarity they exploited a geometric representation. They place nodes in circles whose distance from the origin depends on the birth time while the angular position define the \emph{social identity} of nodes. Initially the network is empty.  At each time $t > 1$, a new node labelled $t$ appears at a random angular position $\theta$ on the circle, with polar coordinates $(r_t, \theta_t)$. In order to implement the trade-off between popularity and similarity, it is assumed that the node evaluates its hyperbolic distance  from other nodes with label $s$ ($s<t$), with coordinates $(r_s, \theta_s)$, by means of the function $x_st = r_s + r_t + ln(\theta_{st}/2)$, where $\theta_{st}$ is the relative angle of the two polar coordinates. The node finally connects to the $m$ nodes with the smallest hyperbolic distance. This approach is interesting, but the mechanism appear rather artificial. In our model, we try to take into account all of these features by considering a simple information dynamics algorithm that allow to us to model not only the preferential attachment mechanism but also the social closeness between individuals.

\section{The model}
\label{model}
 The model is based on a mechanism that emulates people's strategies for acquiring information in social networks, emphasising the local subjective view of an individual and what kind of information the individual can receive when arriving in a new social context.
\begin{figure*}[!htb]
\centering
\subfigure[]
{\includegraphics[width=7.5cm]{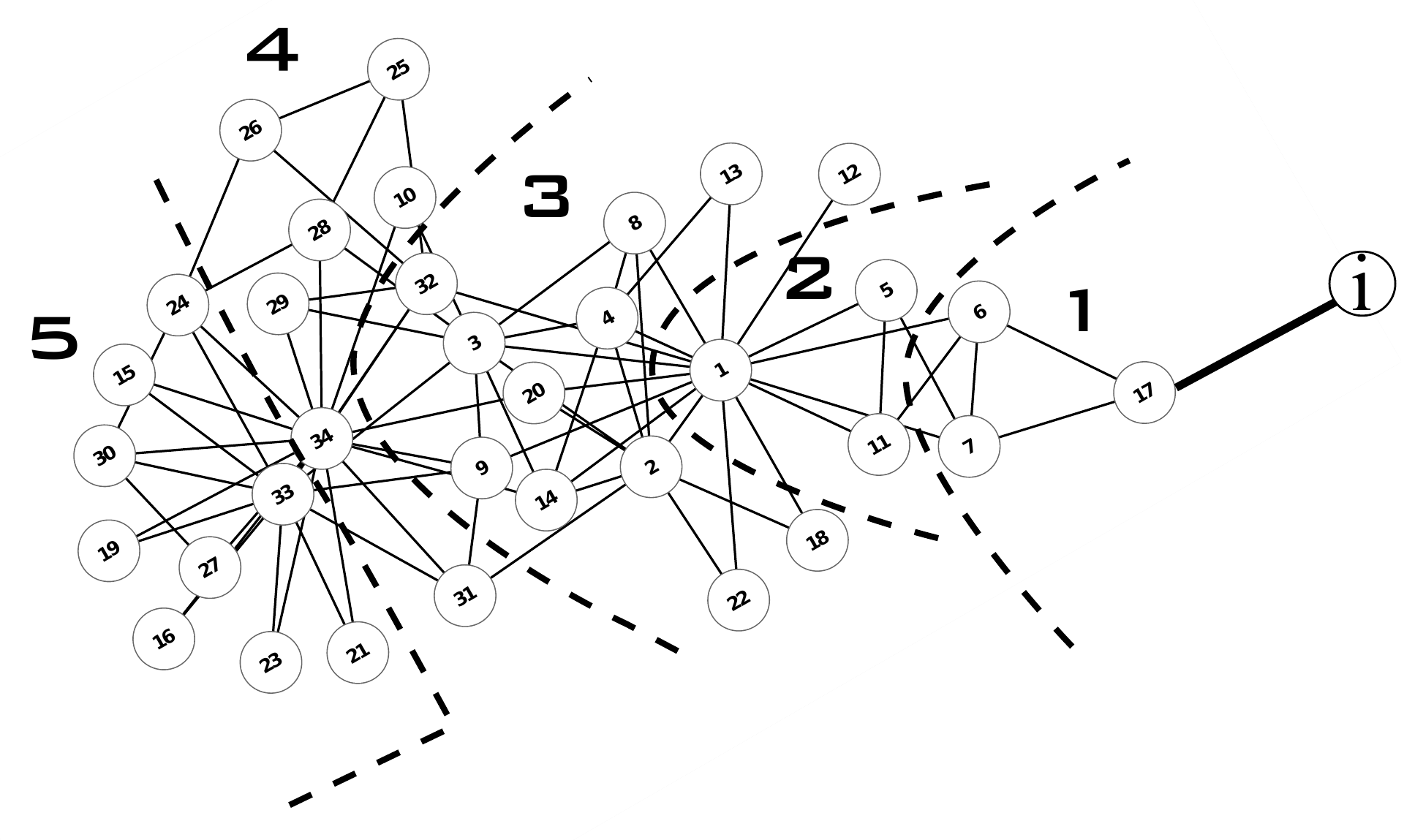}}
\subfigure[]
{\includegraphics[width=8cm]{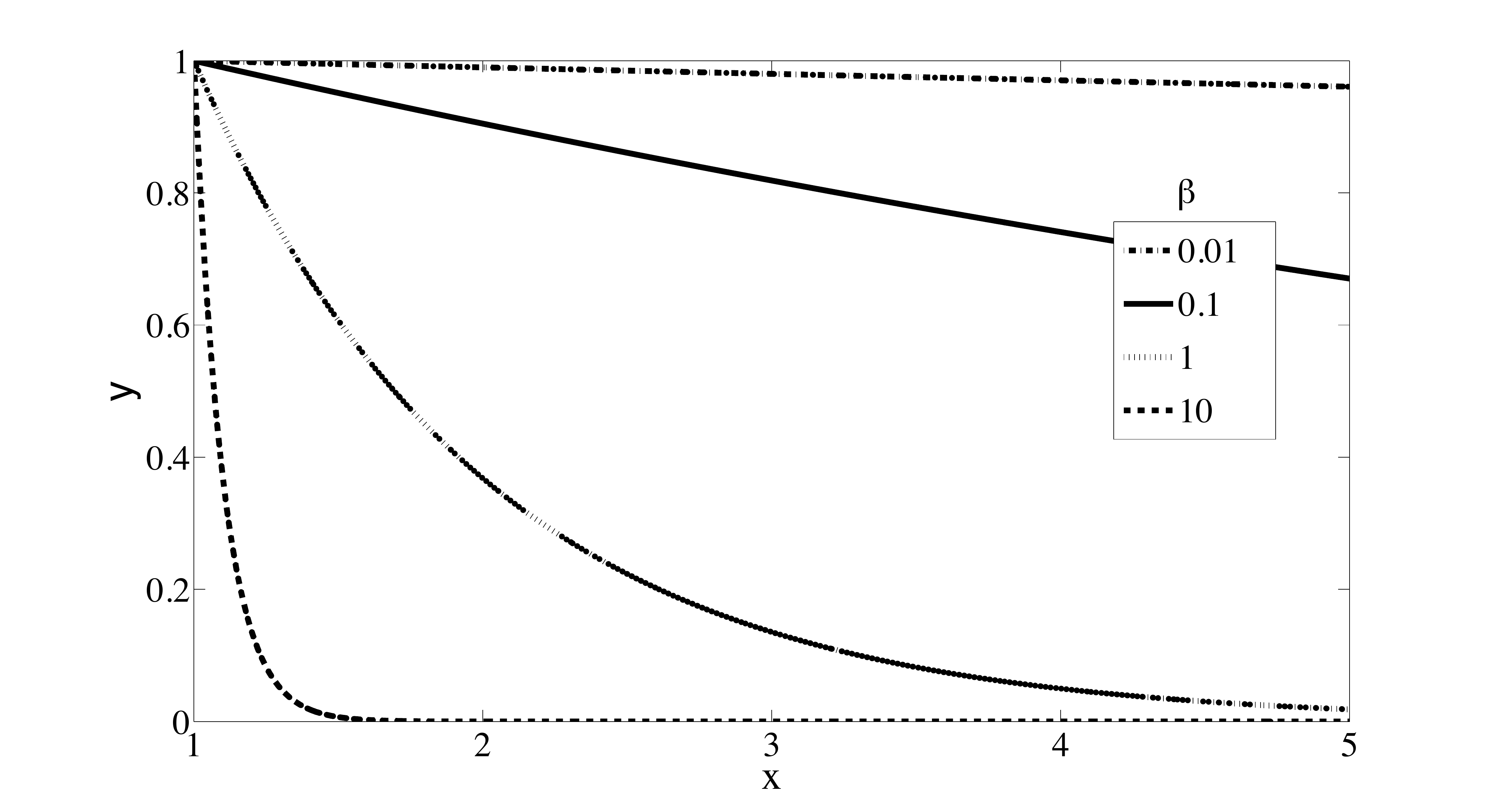}}
\caption{\label{fig:one} (a) Starting configuration of the \emph{Zachary's Karate Club} network~\cite{Zachary}: the new node $i$ links to node $17$. The new node has the local hierarchical representation of the network, labelled by levels $1,2,3,4$ and $5$. With probability $p_1$ it could link to nodes of level $1$, with probability $p_2$ to nodes of level $2$, and so on. (b) Different probability functions derived from Equation \ref{ref:exp} for different values of parameter $\beta$ and $a = 1$. The $y-axis$ represents the probability to join with some nodes in the corresponding level shown on the $x-axis$.}
\end{figure*}

We start from a simple assumption: we suppose that a new individual, or node, arrives in an already structured network and picks information about it. According to this idea we begin, at time $t=0$, with a fully-connected network of $N_0$ vertexes and then, at each time step, a new node is added to  the network. For explaining our idea we describe a simple case, illustrated in  \figurename~\ref{fig:one}~(a), representing a new node $i$ that joins the  \emph{Zachary's Karate Club} network~\cite{Zachary} (assuming that the node $i$ is invited by node $17$). It discovers $n$ ordered levels: the nodes in level $1$ are those adjacent to the connected node (node $17$ in  \figurename~\ref{fig:one}~(a)), nodes in level $2$ are the friends of my friends, and so on.  With different probabilities $p_1>p_2>...>p_n$ it links to nodes in the different levels. This mechanism implements the assumption that the probability of getting new friends in a social context is strictly correlated with the local structure of the network; it is easier that two persons become friends if they a have a friend in common.
In this paper we use a simple exponential function
\begin{equation}
\label{ref:exp}
  y = a e^{-\beta (x-1)},
\end{equation}
where $x$ is the considered level. The  $\beta$ represent the  ``temperature'': the probability of joining farther nodes, and $a$ is a normalization constant. In the \figurename~\ref{fig:one}~(b) we show the function~\eqref{ref:exp} for some values of  $\beta$ with $a=1$.
Assuming $-\beta(x-1)=z$ and $e^z = b^z$, we can express the probability distribution of Eq.\ref{ref:exp} as 
$
\sum\limits_{z=0}^{\infty} ab^z = a\frac{1}{b-1}.$ Setting the previous equation equal to $1$ we obtain $a = 1-b$.

Then,
$
y = (1-b)b^z = \left(1-e^{-\beta}\right) e^{-\beta(x-1)} = P(x).
$
For high values of parameter $\beta$ the probability to join other levels is very low (\textit{e.g.}, the continuous line in  \figurename~\ref{fig:one}~(b)).

The second part of the algorithm allows new nodes to locally elaborate the information about the nodes belonging to a given level. The network is represented by the adjacency matrix  $A_{ij}=1$ $(0)$, $1$ if there is a link between $i$ and $j$, $0$ otherwise. 

Each individual $i$ is characterized by a knowledge vector $S^{(i)}$, representing his knowledge of the world. The knowledge vector $S^{(i)}$ is a probability distribution, assuming that $S^{(i)}_j$ is the probability that individual $i$ knows about the community $j$. It can also be seen as the probability that $i$ belongs to the community "leaded" by $j$, and therefore, $S^{(i)}_j$ is normalized over the index $j$. In order to use a compact notation, we arrange the knowledge vectors for all individuals column by column as $S_{ij} = S^{(i)}_j$, forming a knowledge matrix $S=S(t)$ of the whole network at time $t$.
We initialize the system by setting $S_{ij}(0) = \delta_{ij}$, where $\delta$ is the Kronecker delta, $\delta_{ij}=1$ if $i=j$ and zero otherwise. In other words, at time 0 each node knows only about itself.

The dynamics of the network is given by an alternation of communication and elaboration phases. The communication is implemented as a simple diffusion process, with memory $m$. The memory parameter $m$ allows us to introduce some important features of the human cognitive system, for example that recently acquired information have more relevance than information gained in the past~\cite{Tulving82,Forster84}.

In the communication phase, the state of the system evolves as
\begin{equation}
\label{ref:prima}
  S_{ij}(t+1/2) = m S_{ij}(t) + (1-m) \sum_k A_{ik}S_{kj}(t),
\end{equation}
where $A$ is the adjacency matrix. We assume that nodes talk with each other and we suppose that nodes with high connectivity degree have greater
influence in the process of information's diffusion. This is
due to the fact that during a conversation it is more likely
to know a vertex with high degree instead of one that has
few links. For this reason, the information dynamics is a
function of the adjacency matrix $A$.

The elaboration phase implements elements of fast and frugal heuristics~\cite{Gigerenzer2011}. When people are asked to take a decision, very rarely do they weight all available pieces of information. If there is some aspect that has a higher importance than others, and one item exhibits it, than the decision is taken, otherwise, the second most important factor is considered, etc. In order to implement an adaptive scheme, we exploit a similarity with competition dynamics among species.
\begin{table}[!b]
\caption{\label{tab:result1} Results from the information dynamics algorithm for the different levels in the network shown in \figurename~\ref{fig:one}~(a). $L$ is the number of the level, $n$ is the id of the node and $p$ is the probability to join with the most connected node (bold node).}
\centering
\begin{tabular}{ccc}
\hline
 L&n & p \\
\hline
1& 6, 7 & 0.5-0.5   \\
2& \textbf{1}, 5, 11 & 0.97 \\
3& 2, \textbf{3}, 4, 8, 9, 12, 13, 14, 18, 20, 22, 32 & 0.35  \\
4 & 10, 25, 26, 28, 29, 31, 33 \textbf{34}& 0.67 \\
5 & 15, 16, 19, 21, 23, \textbf{24 }, 27, 30& 0.47\\
\hline
\end{tabular}
\end{table}

If two populations $x$ and $y$ are in competition for a given resource, their total abundance is limited. After normalization, we can assume $x+y=1$, i.e., $x$ and $y$ are the frequency of the two species, and $y=1-x$. The reproduction phase is given by $x' = f(x)$, which we assume to be represented by a power $x'=x^\alpha$. For instance, $\alpha=2$ models birth of individuals of a new generation after binary encounters of individuals belonging to the old generation, with non-overlapping generations \cite{Nicosia09}.

After normalization
$
 x' = \frac{x^\alpha}{x^\alpha + y^\alpha} = \frac{x^\alpha}{x^\alpha + (1-x)^\alpha}.
$
and introducing $z=(1/x)-1$ ($0\le z< \infty$), we get the map
$
 z(t+1) = z^\alpha(t),
$
whose fixed points (for $\alpha > 1$) are 0 and $\infty$ (stable attractors) and $1$ (unstable), which separates the basins of the two attractors. Thus, the initial value of $x$, $x_0$, determines the asymptotic value, for $0\le x < 1/2$ $x(t\rightarrow\infty) = 0$, and  for $1/2< x < 1$ $x(t\rightarrow\infty) = 1$.
By extending to a larger number of components for a probability distribution $S^{(i)}$, the competition dynamics becomes
\begin{equation}
\label{ref:terza}
  S_{ij}(t+1) = \frac{ S_{ij}(t+1/2)^\alpha}{\sum_k  S_{ik}^{\alpha}(t+1/2)},
\end{equation}
and the iteration of this mapping, for $\alpha>1$, leads to a Kronecker delta, corresponding to the largest component. The parameter $\alpha$ allows us to model a ``pruning effect" of the information, which eliminates unnecessary clutter and a clears the way for more information to enter the field of view of the individuals~\cite{brown}. The convergence time depends on the relative differences among the components and therefore, when coupled with the information propagation phase, it can produce interesting behaviours. The model has two free parameters, the memory $m$ and the exponent $\alpha$.

Finally, the probability of making a new link ($P_n$) depends on the joint probabilities of two functions $f(y)$ and $g(S)$:
$
P_n = f(y)\cdot g(S).
$

In summary: (1) we start with $m_0$ nodes ($m_0 \ge 1$); (2) at time $t$ a new node, labelled by $t$, appears in the network; (3) the new node connects with a random node in the network, discovering $n$ levels; with probability $p_1$ given by  Eq~\ref{ref:exp} it joins  the selected level; (4) the new node links with probability $p_2$ given by the Eq~\ref{ref:terza} to the level's nodes. In this way we take into account the social closeness because of the probability to link to nodes in the network depends on the social distance from my \emph{closest} friend \emph{and} the popularity of nodes given by the information dynamics procedure.

\section{Results}
%\begin{figure}[b!]
%\centering
%{\includegraphics[width=7cm]{confronto_BA.pdf}}
%\caption{\label{fig:BA} Cumulative  degree distribution of  networks with $N = 1000$ nodes generated with our model (black dots) (assuming $m=1$ and $y'=1$ for all levels) and the BA model (blue points). }
%\end{figure}
\label{results}
\begin{figure*}[!h]
\centering
\subfigure[]
{\includegraphics[width=7.5cm]{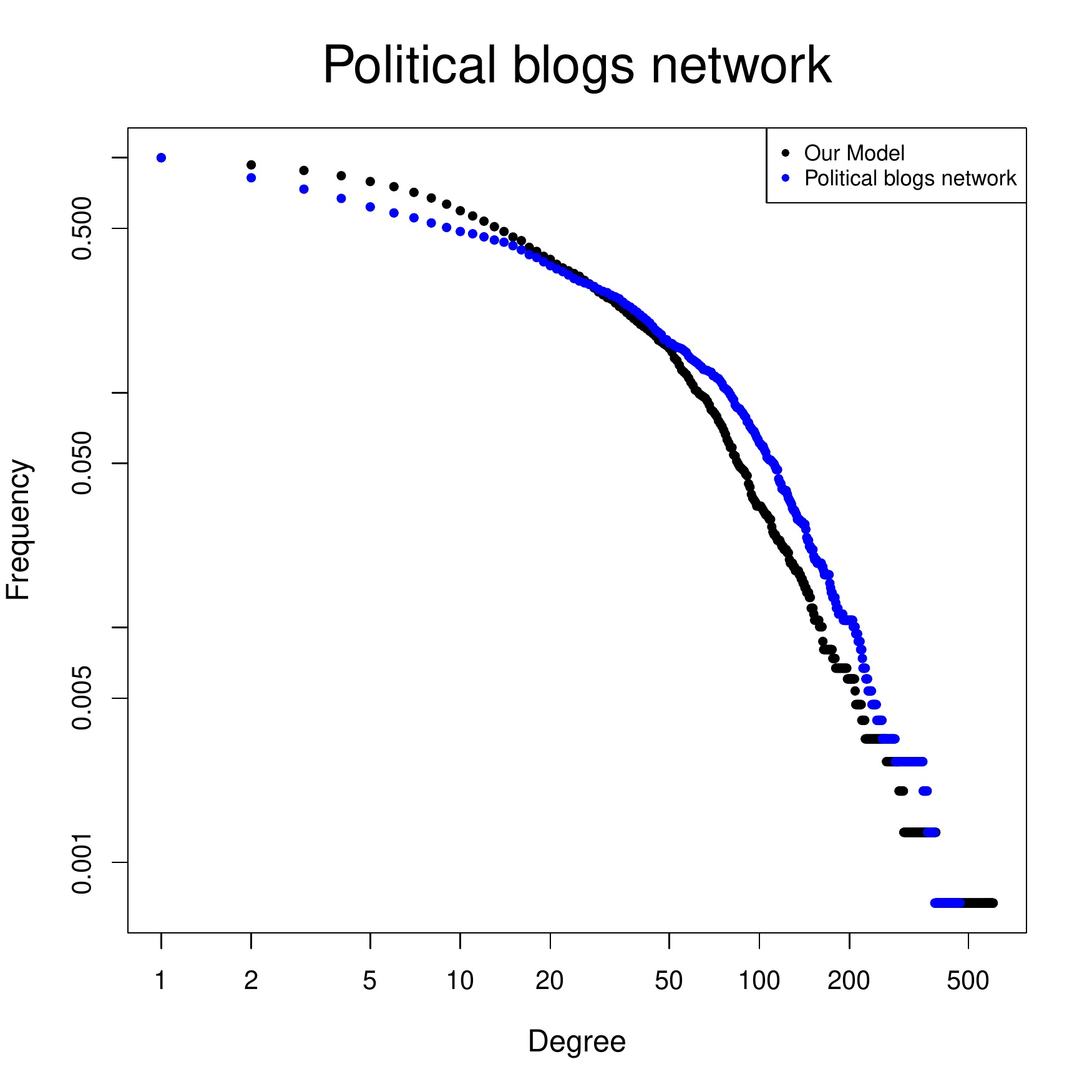}}
\subfigure[]
{\includegraphics[width=7.5cm]{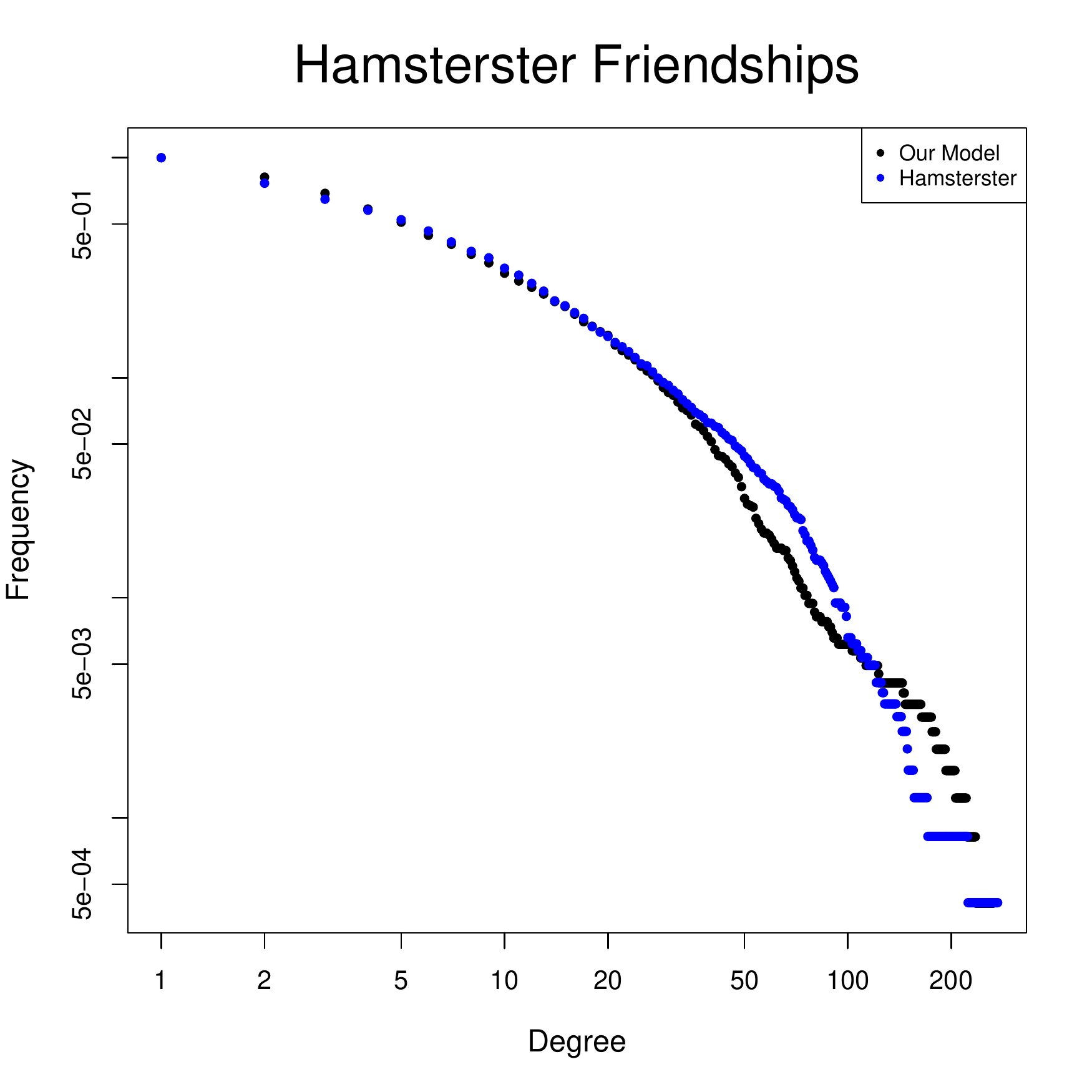}}
\caption{\label{fig:confronto_ok} (a) Cumulative frequency of node degree distributions ($log-log$ scale) of Political blogs network (blue points) and model generated predictions (black points). (b) Cumulative frequency of node degree distributions ($log-log$ scale) of the network of friendships between users of the website hamsterster.com and model generated predictions (black points). Model predictions are averaged over 10 simulation runs. }
\end{figure*}
\begin{figure}[!h]
\centering
{\includegraphics[width=6cm]{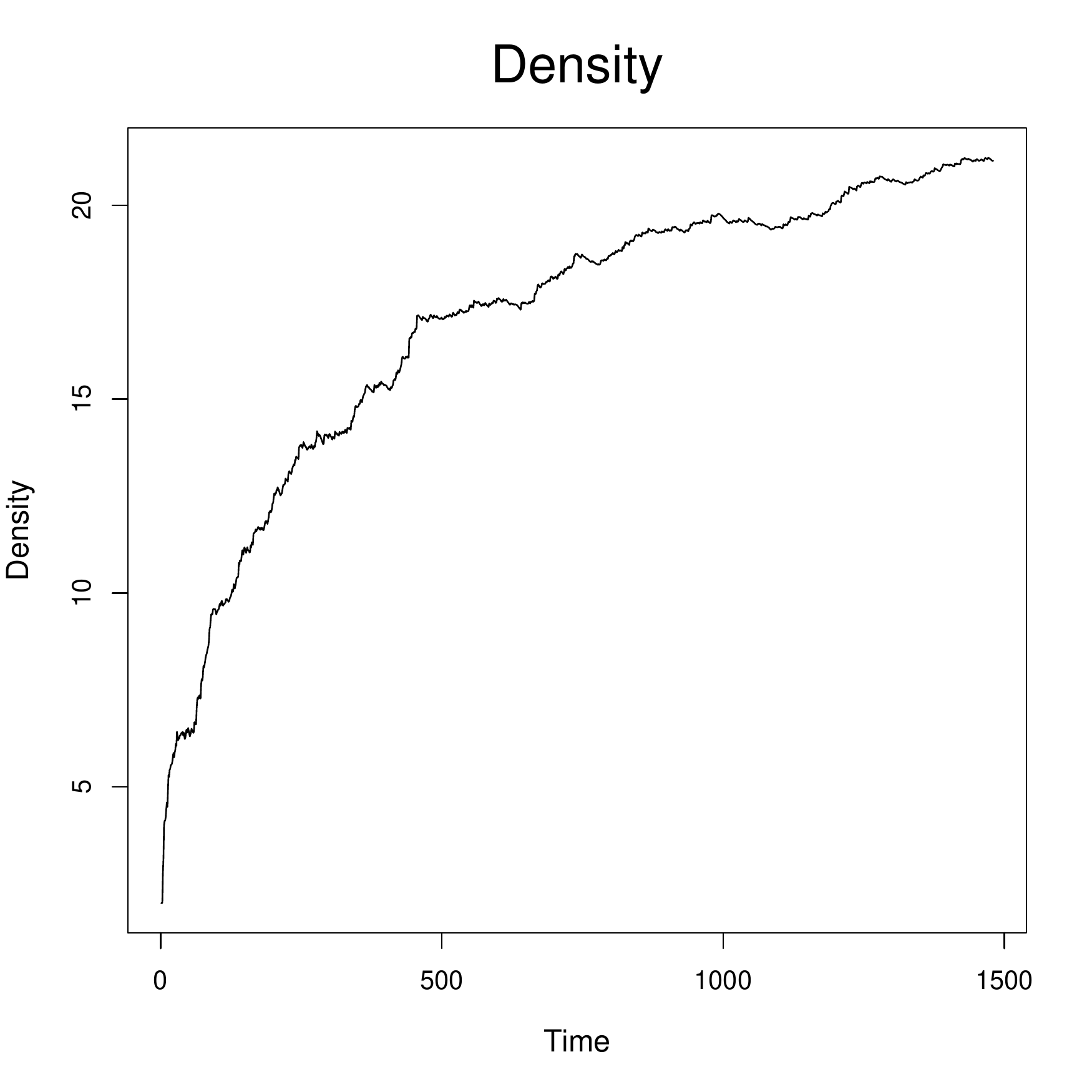}}
\caption{\label{fig:numbers} Simulation of the temporal density evolution of the \emph{Political blogs network}.}
\end{figure}
Results obtained with the information dynamics algorithm, applied to the network represented in \figurename~\ref{fig:one}~(a),  are shown in Table~\ref{tab:result1}. It is more likely that a new node will be connected to a node with a high degree, as is also predicted by preferential attachment.

%From this mechanism it is possible to obtain different kinds of networks. In particular, if we consider that the probability to link to nodes does not depend on the levels, but only on the information dynamics algorithm, we obtain scale-free networks because of the probability to link to hubs is higher than to link to a random node.  \figurename~\ref{fig:BA} shows that the information dynamics algorithm alone can generate scale-free networks. Here we consider $m=1$ and $y' = 1$ for all levels. The scale-free network generated with the BA model was done using the library \emph{igraph} in R~\cite{igraph}.
%Another important feature of real networks is the presence of a strong community structure. Our method allows us to generate networks with different community structures. If we use high values of the parameter $\beta$, the new nodes can join with all levels in the networks. If we use low values of $\beta$, the new node can join only with the closest nodes.

In order to validate our model we compare predictions from the model with two real networks. The model predictions are averages over ten simulation runs.
\begin{table}[!t]
\centering
\caption{\label{tab:net} Statistics of the social networks: $C$ (mean clustering coefficient), $l$ (average path length) and $d$ (diameter of the network).}
\begin{tabular}{cccccc}
 & No.vertices & No.edges& $C$ & $l$ & $d$ \\
\hline
Political blogs & 1490& 19090& 0.24 & 3.39 & 9\\
Simulation & 1490 & & 0.24 & 3.23 & 9 \\
\hline
Hamsterster.com & 2426 & 12534 & 0.09 & 3.54 & 10  \\
Simulation & 2426 & & 0.09 & 3.84 & 11
\end{tabular}
\end{table}
The first social network is the \emph{Political blogs network}. It is a directed network of hyperlinks, with $N = 1490$ nodes, among weblogs on US politics, recorded in 2005 by Adamic and Glance~\cite{pol}.
The degree distribution of the resulting network from a simulation of the model with the same number of nodes is comparable with the real network as shown in \figurename~\ref{fig:confronto_ok}(a).
The mean clustering coefficient of the network generated with our simulations is $C=0.24$, the average path length is $l = 3.23$, and we obtained a network with the same diameter, $d = 9$.
The second social network contains friendships between users of the website hamsterster.com~\cite{hamster}. The degree distribution of the network is shown in  \figurename~\ref{fig:confronto_ok}(b). The cumulative frequency of the degree distribution is very similar to the network generated with the model (black points in  \figurename~\ref{fig:confronto_ok}(b)). The mean clustering coefficient $C = 0.09$, the average path length is $l = 3.84$, and the diameter $d = 11$. Another important feature of the model is that the final structure of the network spontaneously arises without any constraint on the degree of the new node. In fact in our simulations we don't assume any constraints on the number of links that the new node can do: theoretically the new node can link with all the nodes in the network. 
In \figurename~\ref{fig:numbers} we show a simulation of the temporal evolution of network density, defined as the average number of links connected to each node, for the \emph{Political blogs network}. From this result we show that the number of links of incoming node increase over time: in fact the number of link that the new node can do depends on the number of nodes in the network (for instance in a network of $10$ nodes the probability to make a certain number of link is less than in a network of $1000$ nodes). It is a well known result that the density increases in networks that grow over time.

\section{Conclusions}
In this paper, we introduced a new model of growing complex networks. The model is based on the idea that local structure plays a fundamental role in social networks and may be involved also in the growing process of the network itself. Our model reproduces the main features observed in real networks, such as high clustering coefficient, low characteristic path length, strong division in communities and variability of degree distributions.

Following these encouraging results, future work will compare this model with the model proposed by Papadopoulos et al.~\cite{popularity} and validate new results with larger real world networks. Moreover, we plan to derive analytical predictions from the model in order to fix a priori the model's parameters for forecasting some graph's properties (eg. degree distribution, clustering coefficient, diameter, etc.).

\section*{Acknowledgments} This work is financially supported by RECOGNITION Project, a 7th Framework Programme project funded under the FET initiative.

\nocite{*}
\bibliographystyle{eptcs}
\bibliography{bibliogr}
\end{document}